\documentclass[aps,pra, twocolumn,amsmath, longbibliography, amssymb,superscriptaddress]{revtex4-1}
\usepackage{bm}
\usepackage[retainorgcmds]{IEEEtrantools}
\usepackage{graphicx}
\usepackage{mathrsfs}
\usepackage{amsmath}
\usepackage{amssymb}
\usepackage{color}
\usepackage{amsfonts}
\usepackage{times,txfonts}
\usepackage{nicefrac}
\usepackage[colorlinks=true,linkcolor=blue,urlcolor=blue,citecolor=blue,pdfusetitle]{hyperref}
\usepackage{physics}
\usepackage{bbm}

\begin{document}
\title{Charging batteries with quantum squeezing}
\date{\today}
\author{Federico Centrone}
\affiliation{Sorbonne Université, CNRS, LIP6, 4 place Jussieu, F-75005 Paris, France}
\affiliation{Université de Paris, CNRS, IRIF, 8 Place Aurélie Nemours, 75013 Paris, France}
\author{Luca Mancino}
\affiliation{Centre for Theoretical Atomic, Molecular, and Optical Physics, School of Mathematics and Physics, Queens University, Belfast BT7 1NN, United Kingdom}
\author{Mauro  Paternostro}
\affiliation{Centre for Theoretical Atomic, Molecular, and Optical Physics, School of Mathematics and Physics, Queens University, Belfast BT7 1NN, United Kingdom}

\begin{abstract}
We present a scheme for the charging of a quantum battery based on the dynamics of an open quantum system undergoing coherent quantum squeezing and affected by an incoherent squeezed thermal bath. We show that quantum coherence, as instigated by the application of coherent squeezing, are key in the determination of the performance of the charging process, which is efficiency-enhanced at low environmental temperature and under a strong squeezed driving. 
\end{abstract}

\maketitle

\section{Introduction}

The efficient storage and distribution of energy far from its production centers is rapidly becoming one of the economic market drivers and a key technological challenges for the grounding of a sustainable green powered society. Batteries have consequently become a vital technology in modern society and many efforts are being dedicated to improving their performances in terms of capacity, energy density, power and life-time~\cite{Alliance2019}. The boost of nanotechnologies has made the miniaturization of these ``work reservoirs'' a primary matter. As the size of these devices approaches the sub-molecular scale, it becomes reasonable -- and indeed appropriate -- to formulate a quantum mechanical description of their working principles. One of the core questions in this regard is whether non-classical effects can play a useful role in the improvement of the capabilities of energy-storing systems. This has triggered the drawing of theoretical models able to characterize and quantify quantum advantages in terms of non-equilibrium thermodynamical quantities~\cite{Campaioli2018}.

Interesting case-studies of quantum batteries leveraging on discrete~\cite{Campaioli2017, Andolina2019, Caravelli2020} and continuous \cite{FriisHuber} degrees of freedom have been put forward. Needless to say, limiting the study to a unitary charging process severely reduces the application of the models to realistic scenarios. Moreover, the analysis of quantum batteries in the context of open quantum systems may provide additional ways to improve the potentialities of the batteries. In Ref.~\cite{Manzano2016} it was proven for instance that a squeezed thermal reservoir can improve the power and efficiency of a quantum heat engine. Quantum squeezing, which is the effect of reducing the variance of one  quadrature below the uncertainty of the vacuum state, has found many applications in many domains, from quantum optics to quantum technologies~\cite{Lvovsky2015} and grants the possibility of increasing the energy of a bosonic Gaussian system while keeping a null mean value of the fields. 

In this work we study the effects of  squeezing, both as a coherent charging potential and as an incoherent squeezed bath, in the charging of a battery initially prepared in a vacuum state. Our findings reveal that both forms of squeezing efficiently charge the quantum battery, however their simultaneous usage requires to accurately tune the parameters of the potential and the bath in order to enhance the performance of the system and avoid that their effects cancel out.

The remainder of the paper is organized as follows:  in Sec.~\ref{charging} we introduce the notation, formalize the description of the system and characterize the charging scheme; in Sec.~\ref{open} we give details on the open dynamics in terms of its master equation describing the  evolution  of the system coupled to the environment; in Sec.~\ref{energetic} we discuss the thermodynamic quantities of interest and the operational way to measure them, while Sec.~\ref{efficient} is dedicated to the simulation of the charging cycle. We identify the range of parameters of the Hamiltonian and the bath that allow for an improvement of the efficiency of the battery. Finally in Sec.~\ref{powerful} we bound the quantum speed limit of the charging process to compute the power of the storage device. The investigation reported in this paper sheds some light on the role that the quantum coherences enforced by the use fo squeezing have in the charging process of a quantum battery, thus taking the investigation on the potential quantum advantage for the management of energy-storing devices a step closer to a full grasp.

\section{The system and charging cycle} 
\label{charging}
In what follows, we consider the battery as a single-mode harmonic oscillator that is initially prepared in a thermal state. Such initial state is \textit{completely passive}, meaning that it is impossible to extract useful work from it through unitaries. Completely passive states can also be found in literature as Gibbs states or KMS states \cite{Pusz1}. 

We shall consider a fully Gaussian framework where the state of the battery evolves according to a quadratic Hamiltonian in the quadrature operators 
$\hat x=({\hat{a}^{\dagger}+\hat{a}}/){\sqrt 2}$ and $\hat p= i({\hat{a}^{\dagger}-\hat{a}})/{\sqrt 2}$~\cite{GenoniSerafini,SerafiniBook}. 
Since the system is Gaussian, we can translate its description in the phase space: its first moments $\langle \hat{r} \rangle =0$, and its covariance matrix is  $\sigma_{ij}=\langle\{r_i,r_j\}\rangle$, with $\hat{r}=\left(\hat{x}, \hat{p} \right)^T$.
We will consider a thermal state whose first moments are null $\Bar x_{\tau}  = \langle \hat x\rangle_\tau= \Bar p_{\tau}  = \langle \hat p\rangle_\tau= 0$, whereas the covariance matrix of second moments is $\sigma_{\tau}=\coth ({\beta\mu}/{2}) \mathbbm{1}$ with $\beta$ the inverse temperature of the system and $\mu=\hbar\omega$, where $\omega$ is the frequency of the oscillator. The thermal factor $\coth({\beta\mu}/{2})$ is linked to the average number of excitation in the bath as $N=\frac{1}{2}\left[\coth ({\beta\mu}/{2})-1\right]=(e^{\beta\mu}-1)^{-1}$ .

\begin{figure}[b!]
\includegraphics[width=0.85\columnwidth]{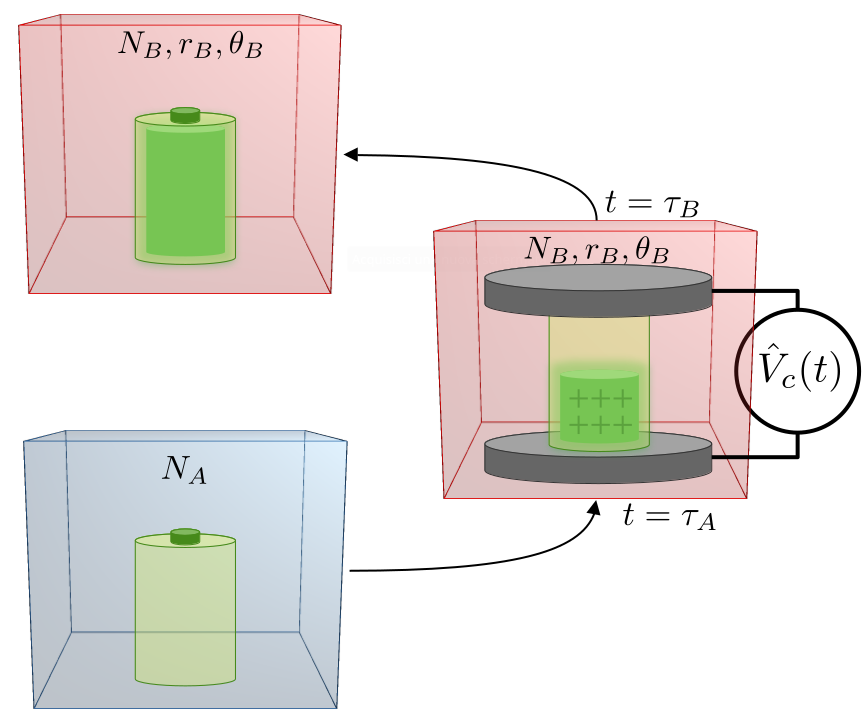}
\caption{Scheme of the charging process proposed in this work. Firstly,  the discharged battery is prepared by letting the system thermalize with a thermal reservoir having an average number of excitation of $N_A$. At time $t=\tau_A$ we turn on the interaction with the charging potential $\hat{V}_c(t)$ and with the squeezed thermal bath that has mean excitation $N_B$, squeezing parameter $r_B$ and squeezing angle $\theta_B$. At time $t=\tau_B$ we turn off the charging potential and the battery is fully charged. }
\end{figure}

First, we aim at implementing a charging operation for a completely passive state $\rho_a$ prepared by letting the battery  thermalize with a reservoir at inverse temperature $\beta_A$, whose  density matrix can be described by $\rho_A=e^{-\beta_A H_0}/Z_A $. The covariance matrix associated to $\rho_A$ is $\sigma_A=\mathbbm{1}(1+2N_A)$, where $N_A$ is the number of excitations within the thermal bath. 

The stroke AB is used to charge the battery. In this stroke, the Hamiltonian of the system is modified by the presence of a charging potential as (from this point on, we assume units such that $\hbar=1$)
\begin{equation}
\hat{H}_{AB}=\hat{H}_0+\hat{V}_c(t),
\end{equation}
where $\hat H_0=\mu(\hat x^2+\hat p^2)/2$ is the Hamiltonian of the oscillator and the charging potential takes the form 
\begin{equation}
\hat{V}_c(t)=-\frac{\lambda}{2}\Sigma(\tau_A,\tau_B)  (\hat{x}\hat{p} + \hat{p}\hat{x})
\end{equation}
with $\Sigma(\tau_A,\tau_B)=\Theta(t-\tau_A)\Theta(\tau_B-t)$ resulting from the composition of two Heaviside step functions, $\Theta(t-\tau_A)$ and $\Theta(\tau_B-t)$ with $\tau_B>\tau_A$, so as to result in a constant in the interval $\tau_{AB}=[\tau_A,\tau_B]$. The charging potential is thus a constant parametric potential of strength $\lambda$ within $\tau_{AB}$, and is null otherwise. 

\subsection{The open dynamics of the quantum battery}
\label{open}
In the following, we aim at characterising the dynamics of an open quantum battery. We assume the system to be weakly coupled to a large environment, whose correlation times are much shorter than the system dynamical time scale and we can always consider them to be uncorrelated, thus allowing us to invoke the validity of the Born-Markov conditions. 
 In such a regime, the dynamics can be described with a Lindblad master equation of the form
    \begin{equation}\label{eq:master}
        \frac{d\hat \rho}{dt}=-i[\hat H_{AB}, \hat \rho]+\sum_{k=1}^m\left( \hat L_k \rho \hat L_k^{\dagger}-\tfrac{1}{2}\{\hat L_k^{\dagger}\hat L_k,\hat\rho  \} \right)
    \end{equation}

    Where $\hat{L}_k$ are the Lindbladian (or jump) operators associated to the non-unitary dynamics. 
    
   Let us introduce a bosonic bath $B$ with quadratures $\mathbf{\hat r}_{\text{bath}}(t)$ satisfying the quantum \textit{white noise} condition
\begin{equation}\label{eq:whitenoise}
    [\mathbf{\hat r}_{\text{bath}}(t),\mathbf{\hat r}_{\text{bath}}(t')]=i\Omega_N \delta(t-t'),
\end{equation}
where $\Omega_N=\Omega^{\oplus^N}$ with $\Omega= i\sigma_y$ 
is the symplectic form (here $\sigma_y$ is the $y$-Pauli matrix).
 Eq.~\eqref{eq:whitenoise} entails the \textit{memoryless} Markovian dynamics, neglecting the correlation of the bath modes at different times. 
    In order to maintain the Gaussian evolution, we can assume a quadratic coupling Hamiltonian $\hat H_C=\mathbf{\hat r}^T C \mathbf{\hat r}_{\text{bath}}^T$ between system and bath. In this situation, the covariance matrix $\mathbf \sigma$ and the first moments $\Bar{\mathbf{r}}$ of the system obey the following diffusive equations
     \begin{equation}\label{eq:diffuse}
         \begin{cases}
         \dot{\Bar{\mathbf{r}}}=A \Bar{\mathbf{r}},  \\
         \dot{\sigma} = A\sigma+\sigma A^T +D,
         \end{cases}
         \end{equation}
         where   the drift matrix $A$ and the diffusion matrix $D$ may be derived from the system hamiltonian $\hat{H}_{AB}$ and its coupling $C$ with the environment.
    
    A key ingredient of our proposal is the squeezed nature of the bath being considered. 
  In this case, we can use the linear response theory as developed in \cite{Mehboudi2019}. The master equation of a system interacting with a squeezed thermal bath is~\cite{Manzano2016} 
 \begin{equation}\label{eq:squeMaster}
        \tfrac{d\hat \rho}{dt}=-\tfrac{i}{\hbar}[\hat H_{AB}, \hat \rho]+\{ \hat L_+ \hat L_+^{\dagger},\hat \rho\}+\hat L_-\hat \rho \hat L_-^{\dagger}-\tfrac{1}{2}\{\hat L_-\hat L_-^{\dagger},\hat \rho  \},
    \end{equation}{}
where the jump operators $\hat L_\pm$ read 
\begin{equation}
\begin{aligned}
   \nonumber \hat L_+&=\sqrt{\tfrac{\Gamma}{2}(N_B+1)}(\hat a \cosh{r_B}+\hat a^{\dagger}\sinh{r_B}e^{i\theta_B}),\\
    \hat L_-&=\sqrt{\tfrac{\Gamma}{2} N_B}(\hat a^{\dagger} \cosh{r_B}+\hat a\sinh{r_B}e^{i\theta_B}).
\end{aligned}
\end{equation}
Here, ${\Gamma}$ is the damping rate, and $N_B=(e^{\beta_B\omega_B}-1)^{-1}$ is the mean number of excitations of a thermal reservoir at  frequency $\omega_B$ and inverse temperature $\beta_B$. and frequency $\omega_B$, $r_B\geq 0$ is the degree of squeezing of the bath and $\theta_B\in[0,2\pi]$ is its phase.  

Since the system's hamiltonian is quadratic in the quadratures, we can rewrite it as  $\hat H_{AB} =\tfrac{1}{2} \mathbf{\hat r}^T H_s \mathbf{\hat r} $, being careful to distinguish the hamiltonian operator $\hat{H}_{AB}$, acting on the Hilbert space of the system, and its hamiltonian matrix $H_s$, mixing the quadratures. The Lindbladian operators, conversely, can be written in the form $\hat L_k=b_k^T \mathbf{\hat r}$. Now, given a master equation such as  Eq.~\eqref{eq:squeMaster}, we can write the drift and diffusion matrix as~\cite{Mehboudi2019}
     \begin{equation}
     \label{AeDdinamiche}
        A=\Omega H_s -\tfrac{1}{2}\text{Im}\left( B B^{\dagger} \right),\quad D=-\Omega \text{Re}\left( B B^{\dagger} \right)\Omega
    \end{equation}
 with $B=(b_1^T, b_2^T,..., b_m^T)\in \mathbb C^{2N\times m}$ taken from the Lindblad operator described above. We can then deduce the form for the matrix $B$ and use it to obtain the drift and diffusion matrices $A$ and $D$.  Plugging these into Eq.~\eqref{eq:diffuse} gives us the dynamical equation for the evolution of the first two moments of our Gaussian system. In what follows, we will focus our study only on vacuum states with null first moments, neglecting the driving of the average value of the quadratures and thus assuming that the quantum state is always fully described by its covariance matrix.

Before describing the dynamics ensuing from Eq.~\eqref{AeDdinamiche}, we shall identify the conditions under which a steady state 
satisfying the stationary equation  
    \begin{equation}\label{eq:steadyState}
        A{\sigma}_{\infty}+{\sigma}_{\infty}A^T+D=0
    \end{equation}
    exists. Criteria for the existence of such a state are provided by the Routh–Hurwitz stability conditions~\cite{RH,DeJK}, which affirms that if $A$ is diagonalizable and the real part of its eigenvalues is negative, then the steady state is stable. When applied to the situation described above, this results in the condition 
    \begin{equation}\label{stabilityCondition}
       \mu^2-\lambda^2-\Gamma^2/4>0.
    \end{equation}{}
This is the condition for the stability of the steady state. Now we want to find a condition on the matrix $D$ in order to enforce the physicality of the dynamics. Imposing the validity of the uncertainty principle for  the bath state covariance matrix ${\sigma}_{\text{bath}}$ we get a \textit{bona fide} diffusive dynamics condition for $D$, which in the single mode case can be reduced as
    \begin{equation}
        \text{Det}[D]\geq \text{Det}[\Omega^T A - A^T\Omega],
    \end{equation}
    which is always satisfied in our case. 
    
\subsection{Energetic considerations} \label{energetic}
The internal energy at time $t$ of a quantum system can be computed as the expectation value of its Hamiltonian $ E=\langle \hat {H} \rangle= \tr[\hat{\rho} \hat{H}]$. As mentioned in Sec.~\ref{charging}, during the charging phase, the Hamiltonian must depend on time in order to change the  energy of the system. However, just outside of the charging period, 
we have $\hat{H}_{AB}(\tau_A^-)=\hat{H}_{AB}(\tau_B^+)=\hat{H}_0$, so that 
\begin{equation}\label{eq:internE}
\begin{aligned}
E=\langle \hat H_0 \rangle=\frac{\mu}{2} \left(\langle  \hat x ^2 \rangle + \langle  \hat p ^2 \rangle \right)=\frac{\mu}{4}\tr[\sigma] 
\end{aligned}
\end{equation}
This expression allows us to derive the internal energy difference between the charged battery at $\tau_B$ and the initial state 
\begin{equation}\label{eq:DeltaE}
    \Delta E_{AB}=E_B-E_A= \frac{\mu}{4}\left(\tr[\sigma_B-\sigma_A] \right),
\end{equation}
where $\sigma_{j}$ is the covariance matrix of the system at time $\tau_j$.

The first law of thermodynamics implies that, for our open quantum system, such energy change is due to two contributions: the work $\Delta W$ done on the system, and the heat $\Delta Q$ exchanged with the environment. These contributions take the form 
\begin{equation}
    \Delta Q={\int_{\tau_A}^{\tau_B} \tr\left[\dot{\hat{ \rho}}(t) \hat{H}(t)\right]\rm{dt}},\, \Delta W={\int_{\tau_A}^{\tau_B} \tr[\hat{ \rho}(t) \dot{\hat{ H}}(t)] \rm{dt}},
\end{equation}
which characterize the work $\Delta W$ spent in order to change the energy of the system  of $\Delta E$ and the amount of heat dissipated to accomplish such result \cite{Vinjanampathy}. For our choice of the Hamiltonian we have
\begin{equation}
\begin{aligned}
    \Delta W_{AB}&= -\frac{\lambda}{2}(\sigma_{B_{12}}-\sigma_{A_{12}}),\\
    \Delta Q_{AB}&=\frac{\mu}{4}\tr[\sigma_B-\sigma_A]+\frac{\lambda}{2}\left(\sigma_{B_{12}}-\sigma_{A_{12}}\right).
\end{aligned}
\end{equation}

The process under consideration is thus not unitary: the thermal bath keeps  draining irreversibly quantum information from the system, increasing its entropy and decreasing its purity until it reaches a non-equilibrium steady state.

None of these quantities, however,  represents the energy available in the battery to perform useful work. This is due to the second principle of thermodynamics which tells us that in a spontaneous process part of the energy is used for increasing the entropy of the system. Therefore we need to consider the Helmoltz free energy defined as
\begin{equation}\label{freeE}
    \Delta F= \Delta E - T \Delta S,
\end{equation}
where $\Delta S$ is the change of the von Neumann entropy $ S = -\tr[ \rho \ln\rho ]$. For Gaussian systems, this can be cast in the form
\begin{equation}
    S=\sum_{i=1}^N\left[\frac{\nu_i+1}{2}\rm{log}\left(\frac{\nu_i+1}{2} \right)-\frac{\nu_i-1}{2}\rm{log}\left(\frac{\nu_i-1}{2} \right)\right],
\end{equation}
where $\nu_i$ is the $i^\text{th}$  symplectic eigenvalue of the covariance matrix $\sigma$. 
The maximum amount of work that the system can perform in a thermodynamic process is given by $-\Delta F$. We will thus use $\Delta F_{BA}$ to characterize the storage capacity of the battery during the discharging process, and the internal energy difference $\Delta E_{AB}$ to quantify the energy required to charge the battery. For the second law, in  a thermodynamic cycle we will always have some irreversible energetic waste so we expect in general $\Delta E_{AB}\geq - \Delta F_{BA}$.

\begin{figure}[t!]
    {\bf (a)}\hskip3.5cm{\bf (b)}\\
    \includegraphics[width=0.52\columnwidth]{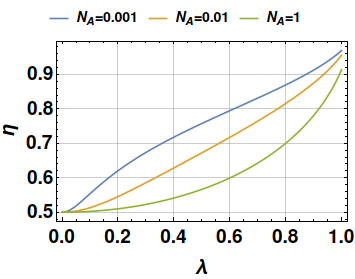}\,\includegraphics[width=0.48\columnwidth]{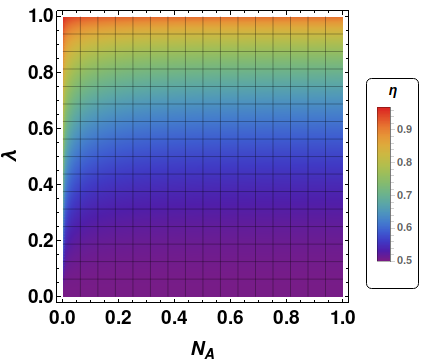}
    \caption{\footnotesize In panel {\bf (a)} we report the trend followed by the efficiency $\eta$ with the temperature of the thermal bath $N_A=N_B$, studied against the charging squeezing $\lambda$. We have taken $\Gamma=\mu=1$ and no squeezing of the bath (i.e. $r_B=0$). Panel {\bf (b)} shows $\eta$ against $\lambda$ and $N_A$ for ?????. }
    \label{fig:eta}
\end{figure}

    In the following we assume $\tau_A=0$ and $\tau_B=+\infty$, so that the charged state of the battery is reached when the system is in the non-equilibrium steady state of the dynamics, and thus $\sigma_B=\sigma_{\infty}$. We will go back to study the dynamical evolution in time when we will discuss the charging power and the quantum speed limits.

\subsection{Efficient charging process}\label{efficient}
The free energy is a function of state that equals zero at thermal equilibrium. As such, it only depends on the initial covariance matrix of the discharged battery $\sigma_A$ and the final state of the charged battery $\sigma_B$ and as a consequence we have that the free energy in the charging stroke equals the free energy in the discharging stroke $\Delta F_{AB}=\Delta F_{BA}$. This means that we do not need to implement the dynamics in the discharging phase in order to characterize the extraction of energy, because this is fully defined by the initial and final state of the charging phase.

In order to compare the performances of the quantum  battery in different dynamical situations, we define the following figure of merit for efficiency
\begin{equation}
    \eta=\frac{\Delta F_{AB}}{\Delta E_{AB}}=1-\frac{\Delta S_{AB}}{\Delta E_{AB}}.
\end{equation}
This corresponds to the ratio between the extractable energy from the battery and the corresponding total internal energy stored in the charged system.

\begin{figure}[t!]
    {\bf (a)}\hskip3.5cm{\bf (b)}\\
    \includegraphics[width=0.53\columnwidth]{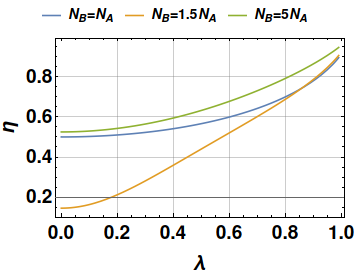}\,\includegraphics[width=0.47\columnwidth]{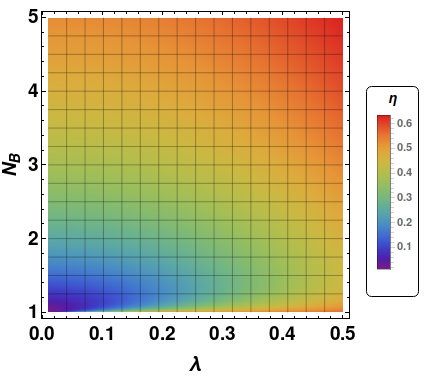}
    \caption{\footnotesize Trend of the efficiency $\eta$ with the temperature of the thermal bath $N_B$ and the charging squeezing $\lambda$. $N_A=\Gamma=\mu=1$ and the squeezing bath is switched off $r_B=0$. }
    \label{fig:etaNB}
\end{figure}

One of the key results of our study is that quantum coherence is a resource for single-mode Gaussian batteries: by increasing the thermal bath temperature $N_A$, and thus decreasing the initial purity of the system, we also decrease its efficiency, despite the fact that the overall available energy is larger. This is shown in Fig.~\ref{fig:eta}, where we report the performance of a single-mode quantum battery system coupled to a single-mode thermal reservoir, setting the squeezing parameter of the bath at $r_b=0$, and taking $N_B=N_A$. Although the dynamical process is non-unitary and the system evolves towards the charged steady-state $\sigma_B$, on average there will be no net heat exchange ($\Delta Q_{AB}=0$) and thus $\Delta E_{AB}=\Delta W_{AB}$. 
Notice that, although both $\Delta F_{AB}$ and $\Delta E_{AB}$ disappear in the limit $\lambda\rightarrow 0$, the efficiency tends to $\eta=1/2$, in this limit. This asymptotic behaviour changes non-trivially if we increase the temperature of the bath $B$, as shown in Fig.~\ref{fig:etaNB}.

\begin{figure*}[t!]
    {\bf (a)}\hskip5cm{\bf (b)}\hskip5cm{\bf (c)}\\
    \includegraphics[width=0.59\columnwidth]{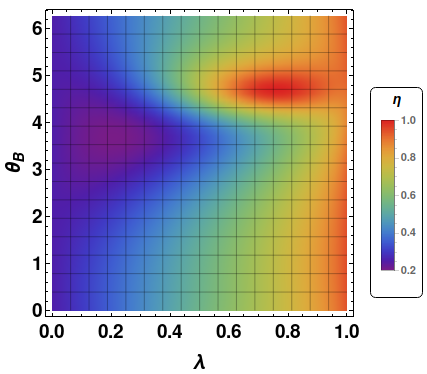}\,\includegraphics[width=0.65\columnwidth]{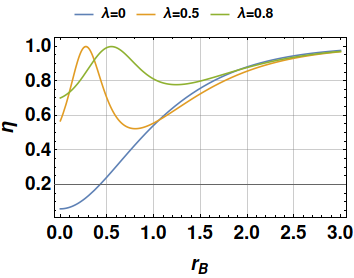}\,\includegraphics[width=0.6\columnwidth]{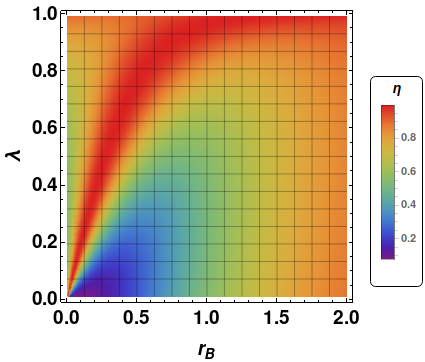}
    \caption{\footnotesize {\bf (a)} Density plot of the efficiency $\eta$ plotted against the charging squeezing $\lambda$ and the phase of the squeezing for the bath $\theta_B$ for $r_B=0.5$. The optimal angle to obtain the maximal efficiency is $\theta_B\sim {3}\pi/2$. {\bf (b)} Trend of the efficiency $\eta$ with the the bath squeezing $r_B$ for various values of the charging squeezing $\lambda$. The phase of the bath squeezing is set to its optimal value, while all the other parameters are as in panel {\bf (a)}. {\bf (c)} Density plot of the efficiency $\eta$ with the  charging squeezing $\lambda$ and the bath squeezing  $r_B$ with the optimal choice of phase for the bath squeezing. In all panels we have taken $N_A=N_B=\Gamma=\mu$.
    }
    \label{fig:etaSQ}
\end{figure*}

We now turn on the interaction with the squeezed bath by setting the parameter $r_b$ to a non-null value. The influence of this type of environment is  complex and the interplay between the various parameters rich. One would expect that, as we increase the squeezing parameters, $\lambda$ and $r_B$, the energy would correspondingly grow. Surprisingly, this is not the case. In fact, the squeezing phase $\theta_B$ of the bath plays a crucial role, and in order to properly charge the battery and improve its efficiency, such parameter should be finely tuned, as it can be appreciated from Fig.~\ref{fig:etaSQ}.

\subsection{Assessment of charging power and temporal considerations}\label{powerful}
We now aim at showing the performance of the average power when charging the battery. We define the average power as
\begin{equation}
    P=\frac{\Delta F_{AB}}{\Delta t_{AB}},
\end{equation}
where $\Delta t_{AB}$ is the average time employed to charge the battery. Although by definition it is required an infinite time for the battery to reach the steady charged state, in the first stages of the dynamics  the system evolves much quicker and then it slows down until it asymptotically reaches the final state, thus $\Delta t_{AB}\neq \tau_B-\tau_A=\infty$. In order to bound the time required to perform the charging we will employ the quantum speed limits geometric formalism. The quantum speed limits bound the minimum velocity $v_{QSL}$ of a system to evolve between a state $\rho$ and a state infinitesimally close $\rho+d\rho$ on the Riemannian manifold formed by the set of density matrices of the Hilbert space of a quantum state. The infinitesimal distance between these states is defined through the Bures metric $ds^2=2[1-\mathcal{F}(\rho,
\rho+d\rho)]$, where $\mathcal{F}$ is the Ulhmann fidelity.

The problem of bounding the minimal Riemannian speed of an infinite dimensional Hilbert space can be challenging \cite{Deffner2017}. However, the limitation to a Gaussian dynamics leads to a critical simplification that allows us to efficiently solve the issue.  In Ref.~\cite{Mancino2020} some of us showed that, for Gaussian states evolving under Gaussian generators, the instantaneous speed of quantum evolution on the Riemannian manifold is 
\begin{equation}
    v^2(t)=\frac{1}{4}\sum_j \frac{\partial_t \nu_j}{\nu_j^2-1}.
\end{equation}
The integral velocity of the system 
between $\tau_A$ and $\tau_B$ is thus
\begin{equation}\label{eq:AVGV}
    V_{AB}=\int_0^{\infty} v(t)dt.
\end{equation}
{This dimensionless quantity embodies the product of the interaction time $\Delta \tau_{AB}$ and the average velocity, which allows us to estimate a lower-bound to the ratio between the average time of the evolution and the interaction time as 
\begin{equation}
    \frac{\Delta t_{AB}}{\Delta \tau_{AB}}=\frac{\Delta s_{AB}}{V_{AB}}.
\end{equation}
}
Here, $\Delta s_{AB}=2[1-\mathcal{F}(\rho_A,\rho_B)]$ is the Bures distance between the passive and charged states. 
Ref.~\cite{Bianchi2015} has provided a closed formula for the evaluation of the Ulhmann fidelity between generic Gaussian states as 
\begin{equation}\label{eq:smFid}
    \mathcal{F}_1^2(\sigma_A,\sigma_B)=\frac{1}{\sqrt{\Delta +\Lambda}-\sqrt{\Lambda}},
\end{equation}
where $\Delta{=}\det[({\sigma_A}+{\sigma_B})/{2}]$ and $\Lambda{=}4\Pi_{j=A,B}\det[({\sigma_j}+{i \Omega})/{2}]$. This gives us all the tools to compute the average power of our single mode Gaussian battery.
\begin{figure}[b!]
{\bf (a)}\hskip3.5cm{\bf (b)}\\
    \includegraphics[width=0.51\columnwidth]{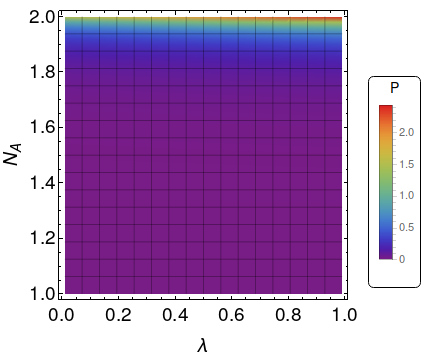}\,\includegraphics[width=0.49\columnwidth]{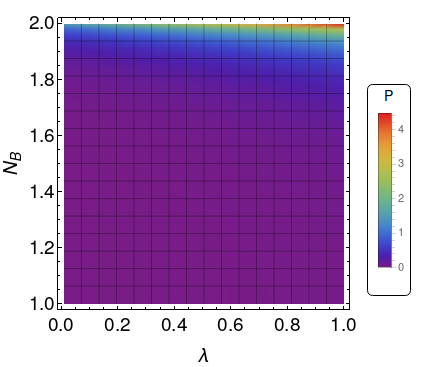}     
    \caption{\footnotesize In panel {\bf (a)} we show the density plot of power $P$ against the mean occupation number $N_A= N_B$ and squeezing $\lambda$. In this simulation, we have taken $\Gamma=\mu$ and a thermal bath with 
    $r_B=0$. Panel {\bf (b)} shows the results of a similar study but for $N_B\geq N_A=\Gamma=\mu$.}
    \label{fig:PLNa}
\end{figure}
Once again, we are going to consider the influence of a simple thermal bath, setting the squeezing parameter of the bath $r_B=0$. We are then going to turn on the charging potential $\hat{V}_c(t)$ with a squeezing strength of $\lambda$. The situation where the two baths $A$ and $B$ have the same temperature $N_A=N_B$ is shown in Fig.~\ref{fig:PLNa} {\bf (a)}.

Differently from the case of the efficiency, the higher temperature, and thus a lower quantum coherence, increases the power of the system. While the increment of $\lambda$ raises the charging power only linearly, the dependence from the initial temperature $N_A$ is actually exponential. Nonetheless,  in the limit of $\lambda \rightarrow 0$ there would be no charging potential and thus no charging power, whereas a pure quantum state at zero temperature can still store energy if $\lambda>0$.


In Fig.~\ref{fig:PLNa} {\bf (b)}, we take $N_A=1$ and let the  bath temperature vary to charge the battery with thermal energy. Even though, the dependence of the charging power from the temperature $N_B$ and the squeezing parameter $\lambda$ is similar to the previous case, the operations performed on the system is conceptually different. In this case, in fact, the energy stored in the battery will increase even if $\lambda=0$.

\begin{figure}[t!]
    {\bf (a)}\hskip3.5cm{\bf (b)}\\
    \includegraphics[width=0.43\columnwidth]{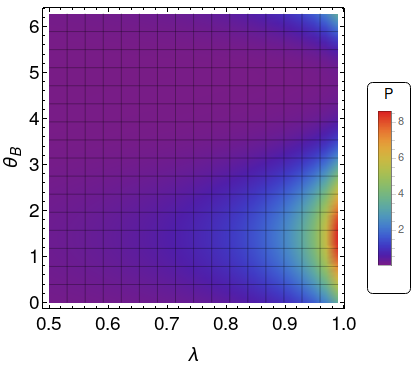}\,\includegraphics[width=0.57\columnwidth]{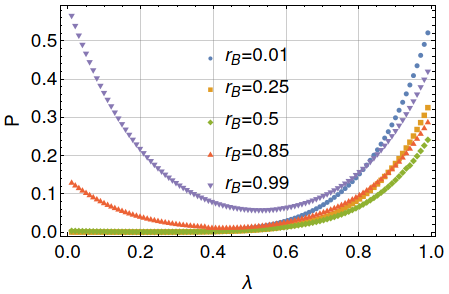}    
    \caption{{\bf (a)} Density plot of power $P$ vs  $\lambda$ and  $\theta_B$. $N_A=N_B=\Gamma=\mu=1$ and the squeezing bath is set to $r_B=0.5$. In this case the optimal value of the angle is $\theta_B\sim\pi/2$. {\bf (b)} Trend of the power $P$ with $\lambda$ for various values of $r_B$. $N_A=N_B=\Gamma=\mu=1$ and the squeezing bath angle is set to its optimal value $\theta_B=\pi/2$.}
    \label{fig:PLT}
\end{figure}

Once again, the situation becomes more complex when we turn on the interaction with the squeezing bath $r_B>0$. In this case, the dynamics strongly depend on the phase of the bath $\theta_B$ and there is an interplay between the two squeezing factors that can be optimized in order to increase the power. Interestingly, the optimal value of $\theta_B$ to maximise the charging power, as shown in Fig.~\ref{fig:PLT}, is different from the optimal value to maximise the efficiency of the discharging (cf. Fig.~\ref{fig:etaSQ}).

This optimal value of $\theta_B$ is used in  Fig.~\ref{fig:PLT} {\bf (b)}, where it is shown the trend of the power $P$ with $\lambda$ for various $r_B>0$. The behaviour of the system is far from being trivial and while we would expect that the power  always increases with $\lambda$ and $r_B$, it is not the case of Fig.~\ref{fig:PLT} {\bf (b)}. 

%

\section{Conclusions}
We have illustrated a scheme for the charging of a quantum battery based on the dynamics of an open harmonic system subjected to the 
effects of a coherent squeezing charging mechanism, and an incoherent squeezed thermal bath. We have characterized the charging process by tracking its efficiency defined in therm of the fraction of extractable energy over the total energy that can be accommodated in the battery itself. We have demonstrated the key role played by quantum coherence in the charging process, whose efficiency is boosted for a low-temperature environment and strong-coherent squeezing driving. 
\appendix

\section{Closed-system dynamics}

\subsection{Channel picture}

We describe the dynamics through a discrete evolution, applying quantum maps to states rather than solving the associated equations of motion in continuous time. In a closed Gaussian system, the evolution of the covariance matrix is described by $\sigma_C=S \sigma_0 S^T$, where $S\in \text{Sp}_{2,\Re}$ (single-mode Gaussian state) is a real symplectic matrix. We will assume that our Gaussian battery has an internal Hamiltonian $\hat H_0$ and that $S$ represents the charging process induced by an external potential applied for some time. Note that we are not considering first moments, which do play a role in the energy of the system, by neglecting linear terms in the charging potential. Without loss of generality, we can apply an Euler (or Bloch-Messiah) decomposition $S=O K$, where $O$ is an orthogonal matrix representing a quadrature rotation, $K$ is diagonal representing single mode squeezing and we disregarded the last orthogonal matrix because it commutes with the identity of the thermal state \cite{GenoniSerafini,SerafiniBook}.

\begin{figure}[h!]
\centering
 {\bf (a)}\hskip4cm{\bf (b)} \\
  \includegraphics[width=0.52\columnwidth]{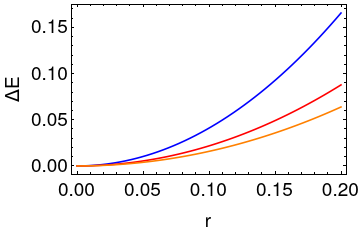}\includegraphics[width=0.48\columnwidth]{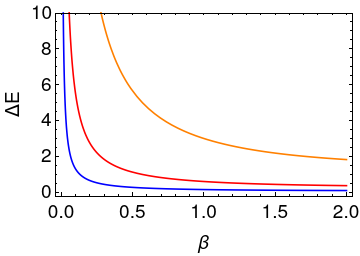}
\caption{Trend of the energy difference (in units of $\mu$) for a battery undergoing closed dynamics. {\bf (a)}: Energy difference $\Delta E$ as a function of the squeezing parameter $r$. From top to bottom curve, we have $\beta=0.5,1$ and $1.5$, respectively.  
{\bf (b)}: Energy difference against the inverse temperature $\beta$ for $r=0.25$, 0.5 and 1 (bottom to top curve, respectively).} 
\label{fig:test}
\end{figure}

We can parametrize as $O=\cos\theta\openone+i\sin\theta\sigma_y$ 
 and $K=\exp[-r\sigma_z]$ 
 with $\sigma_z$ the $z-$Pauli matrix. With this at hand, the most general covariance matrix of a Gaussian state reads
\begin{equation}
    \sigma_C=(1+2N_A)\begin{pmatrix}
    e^{-2r} \cos ^2 \theta + e^{2r}\sin ^2 \theta && \sin({2\theta})\sinh(2r)\\
    \sin(2\theta)\sinh(2r) && e^{2r} \cos ^2 \theta + e^{-2r}\sin ^2 \theta 
    \end{pmatrix}
\end{equation}
This provides information on the charged battery. We can now `unplug' the charger and let the system be driven by its own internal Hamiltonian $\hat H_0$. The energy difference is thus given by Eq.~\eqref{eq:internE}
\begin{equation}
    \Delta E_{AB}=
    \frac{\mu}{2}(1+2N_A)\sinh{(x)}^2
\end{equation}
 and its trend against $r$ and $N_A$ is shown in Fig.~\ref{fig:test}.
Notice that the parameter $\theta$ does not contribute to the energy as rotations are passive transformations. 

The calculation of the ergotropy of the battery would require an optimization of the charging symplectic transformation. However, fixing the bath parameters, the energy difference always grows with the squeezing, so we can  assume  that there is a finite amount of energy or  time to charge the battery. In this case, the work $W$ coincides with $\Delta E$.

\twocolumngrid\
\subsection{Continuous time evolution}

In order to describe the dynamics in time we need to write explicitly the quadratic Hamiltonian of the system, which will be of the form $\hat H=\hat H_0 +\hat V(t)$, where $\hat V(t)=\mu(\hat a^{\dagger} \hat a+\tfrac{1}{2}) - i \lambda (\hat a^{\dagger} \hat a^{\dagger} -\hat a\hat a)$ is applied for a time t and then becomes null.  By using the definition of canonical conjugate variables, as described previously, we can write the full Hamiltonian from time $0$ to $t$ as
\begin{equation}
    \hat H(t)=\frac{\mu}{2}(\hat x^2+ \hat p^2) - \frac{\lambda}{2}(\hat x\hat p+\hat p \hat x)
\end{equation}
We can see that this Hamiltonian is the sum of a harmonic oscillator part, which implements rotations, and a parametric oscillator part, implementing squeezing through a parametric amplification \cite{SerafiniBook}.

\smallskip

Using a bold symbol for the vectorial notation we can write the quadratures as $\mathbf{\hat r}=(\hat x,\hat p)^T$, so that we can write any quadratic Hamiltonian as $\hat H =\tfrac{1}{2} \mathbf{\hat r}^T H_s \mathbf{\hat r} $. In our case $H_s=\mu\openone-\lambda\sigma_x
$ with $\sigma_x$ the $x-$Pauli matrix. 
Generally speaking, this matrix should satisfy the condition $H_s>0$ in order to have a bounded spectrum. When this condition is not satisfied, the system cannot have a steady state and will keep gaining energy indefinitely, which is clearly unphysical. 
{The $H_s>0$ condition implies that $\mu>\lambda$.}

The continuous-time dynamics of a Gaussian system can be described by the Lyapunov equation
 \begin{equation}
     \partial_t \sigma = A \sigma +\sigma A^T
 \end{equation}
 with the drift matrix $A=\Omega H$. 
As we are considering a closed evolution we can take $C=0$. Solving the Lyapunov and computing the energy through Eq.~\eqref{eq:internE} we find
 \begin{equation}\label{tEnergy}
     \Delta E(t)=\frac{\mu\lambda^2(1+2N_A)\sinh^2\left(t\sqrt{\lambda^2-\mu^2}\right)}{\lambda^2-\mu^2}.
     \end{equation}
At first sight, this result appear in contradiction with the stability condition $\mu>\lambda$. However we can verify that the solution exists and is continuous for all real values of $\lambda$ and $\mu$. In fact
\begin{equation}\Delta E(t)=
     \begin{cases}
          \frac{\mu\lambda^2(1+2N_A)\sin^2\left(t\sqrt{\mu^2-\lambda^2}\right)}{\mu^2-\lambda^2}~\text{for}~ \mu>\lambda,\\
           \mu\lambda^2\left(1+2N_A\right)t^2~\text{for}~\mu \sim \lambda.
     \end{cases}
\end{equation}
The last expression has been found using the Taylor expansion of $\sin(t \sqrt{\lambda^2-\mu^2})$.
     \begin{figure}[h]
  \includegraphics[width=\linewidth]{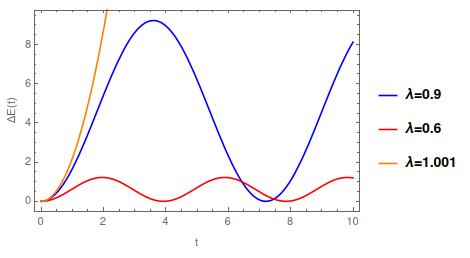}
  \caption{Trend of energy in time}
  \label{fig:tEnergy}
\end{figure}
In particular we can see how the energy is bounded only inside the stability condition, as predicted by the theory.

\section{Multimode system}

The study on single-mode Gaussian batteries can be readily extended to the multimode case with little differences. 
Eq.~\eqref{eq:smFid} for the fidelity of single mode Gaussian states, needs to be generalized to the multimode case using the expression~\cite{Bianchi2015}
\begin{equation}\label{eq:Gfid}
    \mathcal{F}^2(\sigma_A,\sigma_B)=\frac{F_{\text{tot}}}{\sqrt[\leftroot{-2}\uproot{2}4]{\det (\sigma_A+\sigma_B)}
}
\end{equation}
with
\begin{equation}
    F^4_{\text{tot}}=\det \left[ 2\left(\sqrt{\mathbbm{1}+\frac{(\sigma_\text{aux}\Omega )^{-2}}{4}} +\mathbbm{1} \right) \sigma_{\text{aux}} \right]
\end{equation}
and 
$\sigma_{\text{aux}}=\Omega^T(\sigma_A/2+\sigma_B/2)^{-1}(\Omega/4+\sigma_A\Omega\sigma_B/4)$.

All the quantities of interest can now be computed following the analysis of the previous section. One can prove, however, that the multimode case can be reduced to the analysis of a product of single modes system and environments. This result comes from the fact that the we only need the symplectic decomposition of the system and environment joint covariance matrix to derive the thermodynamically relevant quantities of our study,  and all passive elements that can mix modes, although they can deeply influence the dynamics, would not affect the thermodynamics.

\acknowledgements

We acknowledge support from the ANR-18-IDEX-0001, IdEx Université de Paris, the H2020-FETOPEN-2018-2020 project TEQ (grant nr. 766900), the DfE-SFI Investigator Programme (grant 15/IA/2864), COST Action CA15220, the Royal Society Wolfson Research Fellowship (RSWF\textbackslash R3\textbackslash183013), the Leverhulme Trust Research Project Grant (grant nr.~RGP-2018-266), the UK EPSRC (grant nr.~EP/T028106/1).


\begin{thebibliography}{99}

\bibitem{Alliance2019} Global Battery Alliance. \textit{A Vision for a Sustainable Battery Value Chain in 2030: Unlocking the Full Potential to Power Sustainable Development and Climate Change Mitigation}. In World Economic Forum. (2019)

\bibitem{Campaioli2018} F. Campaioli, F. A. Pollock, and S. Vinjanampathy, \textit{Quantum batteries}. In Thermodynamics in the Quantum Regime, F. Binder, L. A. Correa, C. Gogolin, J. Anders, and G. Adesso Eds. (Springer, Cham, 2018)

\bibitem{Campaioli2017} F. Campaioli, F. A. Pollock, F. C. Binder, L. Céleri, J. Goold, S. Vinjanampathy, and K. Modi, \textit{ Enhancing the charging power of quantum batteries}, Phys. Rev. Lett. {\bf 118}, 150601 (2017).


\bibitem{Andolina2019} G. M. Andolina, M. Keck, A. Mari, M. Campisi, V. Giovannetti, and M. Polini, \textit{Extractable work, the role of correlations, and asymptotic freedom in quantum batteries}, Phys. Rev. Lett. {\bf 122}, 047702 (2019).

\bibitem{Caravelli2020} F. Caravelli, G. Coulter-De Wit, L. P. García-Pintos, and A. Hamma, \textit{ Random quantum batteries}, Phys. Rev. Research {\bf 2}, 023095 (2020).

\bibitem{FriisHuber} N. Friis and M. Huber, {\it Precision and Work Fluctuations in Gaussian Battery Charging}, Quantum {\bf 2}, 61 (2018).

\bibitem{Manzano2016} G. Manzano, F. Galve, R. Zambrini, and J. M. R. Parrondo, {\it Entropy production and thermodynamic power of the squeezed thermal reservoir}, Phys. Rev. E {\bf 93}, 052120 (2016).


\bibitem{Lvovsky2015} A. I. Lvovsky, \textit{Squeezed light}. Photonics: Scientific Foundations, Technology and Applications {\bf 1}, 121 (2015).

\bibitem{Pusz1} W. Pusz and S. L. Woronowicz, {\it Passive states and KMS states for general quantum systems}, Comm. Math. Phys. {\bf 58}, 273 (1978).


\bibitem{GenoniSerafini} M. G. Genoni, L. Lami, and A. Serafini, {\it Conditional and unconditional Gaussian quantum dynamics}, Contemp. Phys. {\bf 57}, 331 (2016).

\bibitem{SerafiniBook} A. Serafini, {\it Quantum Continuous Variables} (CRC Press, Taylor \& Francis Group, 2017).

\bibitem{Mehboudi2019} M. Mehboudi, J. M. R. Parrondo, A. Ac\'{i}n, {\it Linear response theory for quantum Gaussian processes}, New J. Phys. {\bf 21}, 083036 (2019).

\bibitem{DeJK} E. X. DeJesus and C. Kaufman, {\it Routh-Hurwitz criterion in the examination of eigenvalues of a system of nonlinear ordinary differential equations}, Phys. Rev. A {\bf 35}, 5288 (1987).

\bibitem{RH} A. Hurwitz, in Selected Papers on Mathematical Trends in Control Theory (R. Bellman and R. Kalaba eds., Dover, New York, 1964).


\bibitem{Vinjanampathy} S. Vinjanampathy and J. Anders, {\it Quantum Thermodynamics}, Contemp. Phys. {\bf 57}, 545 (2016).

\bibitem{Deffner2017} S. Deffner, {\it Geometric quantum speed limits: a case for Wigner phase space}, New J. Phys. {\bf 19}, 103018 (2017).

\bibitem{Mancino2020} L. Mancino, M. G. Genoni, M. Barbieri,  and M. Paternostro, {\it  Nonequilibrium readiness and precision of Gaussian quantum thermometers}, Phys. Rev. Research {\bf 2}, 033498 (2020).

\bibitem{Bianchi2015} L. Banchi, S. L. Braunstein, and S. Pirandola, {\it Quantum fidelity for arbitrary Gaussian states}, Phys. Rev. Lett. {\bf 115}, 260501 (2015).

\bibitem{Imparato2020} K. V. Hovhannisyan, F. Barra, A. Imparato, {\it Charging by thermalization}, arXiv:2001.07696 (2020).









\end{thebibliography}
\end{document}